\documentclass{article}
\usepackage{spconf,graphicx}
\usepackage{amsmath,amsfonts,amssymb}
\usepackage[utf8]{inputenc}
\graphicspath{ {figures/} }\usepackage[colorlinks=true, allcolors=blue]{hyperref}
\usepackage{algpseudocode}
\usepackage{algorithm}
\def\UrlBreaks{\do\/\do_}

\title{Joint Estimation of the Room Geometry and Modes with Compressed Sensing}
\name{Helena Peić Tukuljac, Thach Pham Vu, Hervé Lissek, Pierre Vandergheynst\thanks{This work was supported by the Swiss National Science Foundation under Grant No. 200021\texttt{\char`_}169360 for the project \textit{“Compressive Sensing applied to the Characterization and the Control of Room Acoustics”}. }}
\address{School of Computer and Communication Sciences\\
Ecole Polytechnique Fedérale de Lausanne (EPFL), 
CH-1015 Lausanne, Switzerland\\
\{helena.peictukuljac, thach.phamvu, herve.lissek, pierre.vandergheynst\}@epfl.ch}

\begin{document}
%
\maketitle
\begin{abstract}
Acoustical behavior of a room for a given position of microphone and sound source is usually described using the room impulse response. If we rely on the standard uniform sampling, the estimation of room impulse response for arbitrary positions in the room requires a large number of measurements. In order to lower the required sampling rate, some solutions have emerged that exploit the sparse representation of the room wavefield in the terms of plane waves in the low-frequency domain. The plane wave representation has a simple form in rectangular rooms. In our solution, we observe the basic axial modes of the wave vector grid for extraction of the room geometry and then we propagate the knowledge to higher order modes out of the low-pass version of the measurements. Estimation of the approximate structure of the $k$-space should lead to the reduction in the terms of number of required measurements and in the increase of the speed of the reconstruction without great losses of quality.
\end{abstract}
\section{INTRODUCTION}
\label{sec:intro}

In 2006 Ajdler et al. \cite{PAF} have defined the Plenacoustic function (PAF) as the function that contains the room impulse responses (RIRs) for all the possible pairs of microphone and source positions in a room with the given acoustical properties. Without having any prior knowledge involved, it is extremely hard to estimate the PAF. As shown by Moiola et al. \cite{sumplanewaves} the acoustical behavior of the room can be described by a discrete sum of plane waves that can exist inside a given room which are tightly related to the resonant frequencies.
This plane wave approximation holds for any star-convex room and is independent of boundary conditions, domain of propagation, type of the source or proximity to the source or the walls \cite{Low}.

Sparse plane wave approximation in the low frequency domain introduces an assumption required for sparse analysis of room's complex wavefield which further opens the door to compressed sensing \cite{DonohoCS, CandesCS}. Mignot et al. \cite{Low} have started the trend of the sparse modal analysis. They have designed a greedy approach which uses space decomposition based on iterative alternating projections for the estimation of the wave number and wave vectors that fully determine the acoustical behaviour of the given room. Due to the high dimensionality of data acquired by microphones, greedy methods such as Simultaneous Orthogonal Matching Pursuit (SOMP) \cite{SOMP} (simultaneous, since we are fitting measurements from multiple microphones at once) have shown better performance than the relaxation of the minimization of $\ell_0$ norm \cite{SparseCompare}.

Our solution focuses on the structured sparsity of the plane wave representation for the reconstruction of parameters of the Room Transfer Function (RTF). In literature, sparse plane wave representation has been used not only for the representation of the wavefield in a room in low frequency domain, but also for efficient storage of highly correlated recordings of dense microphone arrays \cite{WaveRepresentation}. Besides sparse plane wave representation an interesting sparse approach to the estimation of RTF is a recent approach with orthonormal basis functions based on infinite impulse response filters (IIR) \cite{OrthonormalBasisFunctions}. Though not exploring plane wave sparsity, the solution relying on the weighted spatio-temporal representation \cite{SpatioTemporal} also gives promising room impulse response interpolations.

On the other hand, the solutions for estimating the shape of the room usually rely on knowing the location of early reflections \cite{MirandaShape}, \cite{IvanShape}, but finding the true reflections within an echogram is not a trivial problem and is still an open research question.

\section{PROBLEM SETUP}
When sampling sound we need to take into account two types of possible aliasings: temporal and spatial. Depending on the highest frequency that we want to capture $f_c$, we define our temporal sampling step $\Delta t$ in such a way that the sampling frequency satisfies $f_s=\frac{1}{\Delta t} > 2f_c$ \cite{Nyquist}. Once the temporal sampling step is fixed, we determine the appropriate sampling step in space either by the limits imposed by the Courant–Friedrichs–Lewy condition \cite{CFL} for finite difference time domain (FDTD) schemes, or by a contemporary view of the problem observed through the sampling of the PAF \cite{PAF}. 

The support of the spectrum of the PAF $\hat{p}(\omega,\varphi_x,\varphi_y,\varphi_z)$, where $\omega=\frac{2\pi}{\Delta t}$ is the temporal angular sampling frequency $[rad.s^{-1}]$ and $\varphi_i=\frac{2\pi}{\Delta i}$ is the spatial angular frequency $[rad.m^{-1}]$ over each of the $i^{\textrm{th}}$ observed axis, lays inside a hypercone:
$\varphi_x^2+\varphi_y^2+\varphi_z^2 \leq \frac{\omega^2}{c^2}$, where $c$ is the celerity of sound. This gives the following condition for the sampling step over each of the axes:
$\Delta i < \frac{\pi c} {\omega_c}, \; \forall i \in \{x,y,z\}$.
The following question emerges: can we acquire the targeted information at lower sampling rates, both in time and in space, by exploiting the underlying structure of the data without introducing significant losses?

\subsection{Plane wave representation of wavefield}

\begin{figure}

\begin{minipage}[b]{1.0\linewidth}
 \includegraphics[width=\columnwidth]{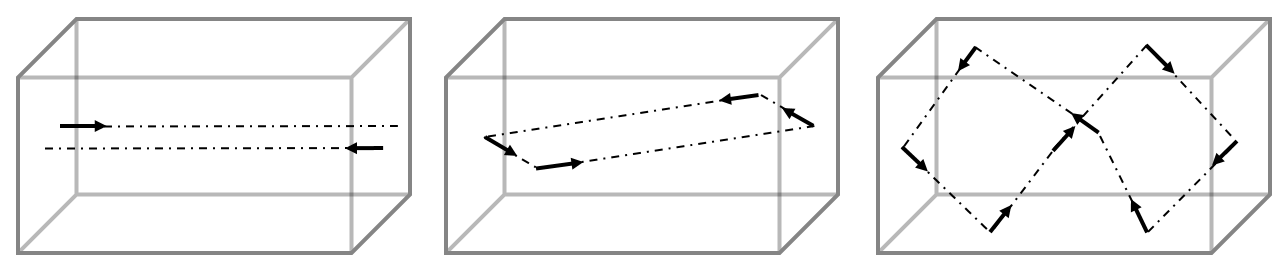}
  \centerline{(a) Plane waves inside a rectangular room.}\medskip
\end{minipage}
\begin{minipage}[b]{1.0\linewidth}
\includegraphics[width=\columnwidth,height=4cm]{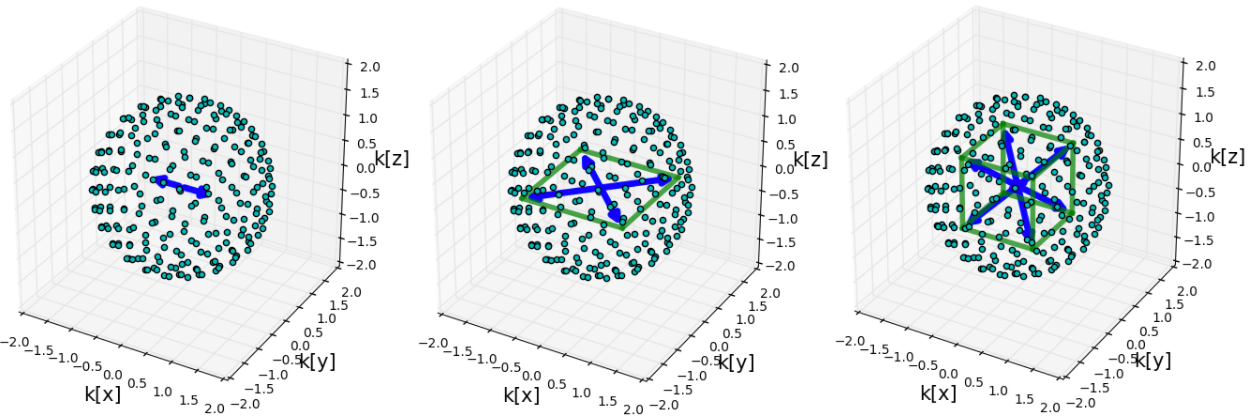}
  \centerline{(b) Wave vectors in the search space.}\medskip
\end{minipage}
\hfill

\caption[width=\textwidth]
{ \label{fig:room_modes} Plane wave types and corresponding structured sparsity of wave vectors. In theory these vectors form a parallelepiped inscribed into a sphere with radius $\frac{\omega_n}{c}$, resulting in structured sparsity. From left to right: $x$-axial mode, $xy$-tangential mode and oblique mode.}
\end{figure}
Acoustic propagation is governed by the wave equation:
\begin{equation}
\Delta p(t,\mathbf{X})-\frac{1}{c^2}\frac{\partial^2}{\partial^2 t}p(t,\mathbf{X})=0.
\end{equation}
Solution of the wave equation can be approximated in the low frequency domain as a discrete sum of damped complex harmonics \cite{sumplanewaves}:
\begin{equation}
p(t,\mathbf{X})=\sum_{q \in I} A_q \phi_q (\mathbf{X}) g_q(t),
\end{equation}
where $I \subset \mathbb{Z}$, $\phi_q$ represents the spatial dependency of mode shape whose shape is illustrated in \cite{MySPIE} and $g_q$ is corresponding time evolution of the mode. Temporal functions are orthogonal. This expression emphasizes the separability of the analysis and the estimation of the temporal and spatial parameters, which can greatly reduce the computational complexity of the parameter analysis \cite{Low}.

Temporal functions take the form of $g_q(t)=e^{j k_q c t}$, where $k_q=\frac{\omega_q-j\xi_q}{c}$ is the \textit{wave number} of the $q^{\textrm{th}}$ room mode. $\omega_q$ is the resonant frequency and $\xi_q < 0$ is the corresponding damping factor. On the other hand, in the spatial functions, the room modes can be decomposed as a sum of plane waves: $\phi_q(\mathbf{X})\approx \sum_{r=1}^R a_{q,r}e^{j \mathbf{k}_{q,r} \cdot \mathbf{X}}$, where $\mathbf{k}_{q,r}$ is the $r^{\textrm{th}}$ \textit{wave vector} of the $q^{\textrm{th}}$ mode and $R = 8$ for a rectangular room case. In Figure \ref{fig:room_modes}. we see an example of all types of plane waves in a rectangular room: axial, tangential and oblique determined by the wave vectors. In the case of a room with \textit{low damping}, the length of the wavevectors can be approximated by the real part of the corresponding wavenumber: $\lVert \mathbf{k}_{q,r} \rVert = |k_q|$, since in that case $k_q \approx \frac{\omega_q}{c}$. This gives us an intuition for the spherical vector search which will be explained more in detail later. For a rectangular room the wave vectors are on the vertices of a parallelepiped inscribed into the sphere. Through modal decomposition (3) and plane waves approximation, the final form of the RIR is composed of the modal wave numbers $k_q$'s, the corresponding wave vectors $\mathbf{k}_{q,r}$'s and their expansion coefficients $\alpha_q$'s:

\begin{equation}
p(t,\mathbf{X})=\sum_{q,r} \alpha_q e^{j(k_q c t+\mathbf{k}_{q,r} \cdot \mathbf{X})}.
\end{equation}

As in the theory of modal decomposition \cite{RoomAcoustics, sumplanewaves}, we will focus on the coupling of the pressure field with the standing waves at room's resonant frequencies.

\subsection{Periodicity of the wave vector grid}
\begin{figure}
\includegraphics[width=\columnwidth]{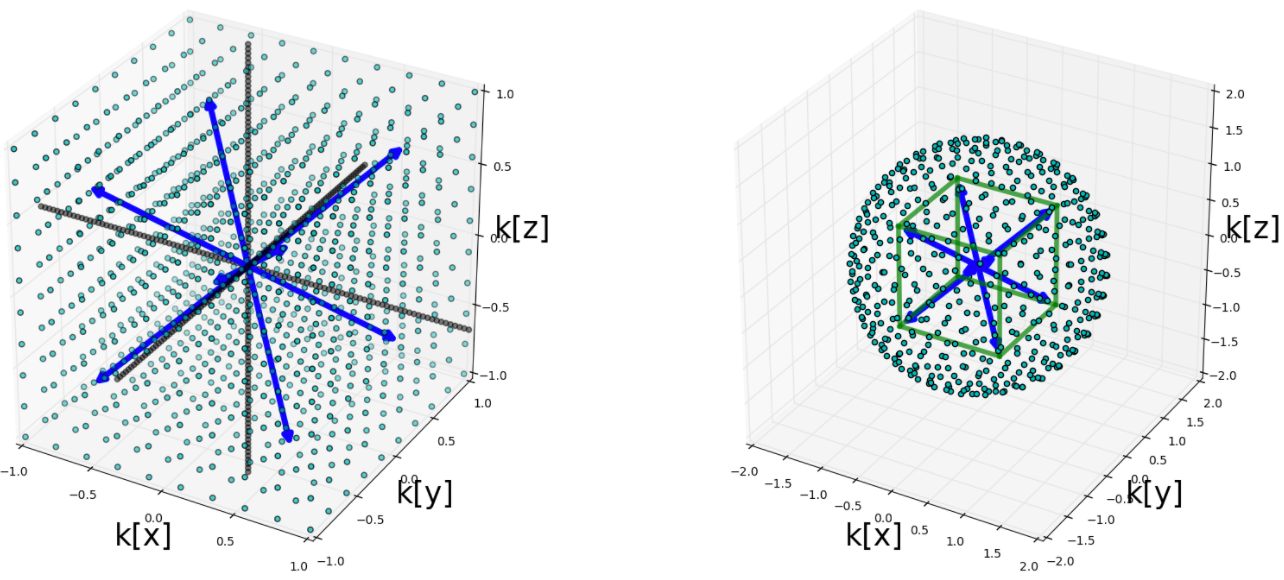}
\caption[width=\textwidth]
{ \label{fig:plane_waves} The left hand side shows the periodicity of the wave vector grid with respect to $\mathbf{k}=[\pm k_x, \pm k_y, \pm k_z]$ with period over the axes equal to: $\frac{\pi}{L_x}$, $\frac{\pi}{L_y}$ and $\frac{\pi}{L_z}$. Here we see an example of an oblique wave vector. The right hand side shows the search space on our uniformly sampled sphere.}
\end{figure}
In our solution we will be focusing only on the rectangular rooms with the regular wave vector grid (regular eigenvalue lattices in the wave vector space) \cite{RoomAcoustics} as shown in Figure \ref{fig:plane_waves}. The \textit{k-space} is an array of numbers representing spatial frequencies.  According to theory, as long as we know the periodicity of the grid over each of the axes, it will provide us the knowledge on the room geometry as well as the values of the wave vectors of higher order. So the goal of our approach is the estimation of these three periods along each of the axes. Under the assumptions that the room is lightly damped, the 3 fundamental axial modes can be used as a basis to find all higher order modes. This will reduce the cutoff frequency of the analyzed data, which further reduces the density of the required grid of microphones, due to the dependencies between the temporal and spatial sampling as shown earlier.

\section{Parameter estimation with partial compressed sensing for structured data}

In 1985, Richardson et al. \cite{CurveFitting} have proposed a curve fitting algorithm allowing the reconstruction of the RTF curve from discrete measurements using room mode shaped functions as basic fitting elements. For different positions of the microphones/sound sources across the room, some parameters stay the same - \textit{common parameters}: eigenfrequencies which depend on the room geometry, and the room mode damping which depends on the damping of the walls. The attenuation and the phase of the room modes are position dependent parameters - \textit{specific parameters} which are expressed by different expansion coefficients and different spatial coordinates.

\subsection{Acoustical properties of rectangular rooms}
There are two key points for our parameter estimation procedure: how many room modes $N$ do we expect up to a given cutoff frequency $f_c$ and what are their approximate resonant frequencies $\omega_n$?
These are dependent on the room shape and size \cite{RoomAcoustics}. For a rectangular room of size $L_x \times L_y \times L_z$ angular eigenfrequencies are given by the expression: $\omega_n = \pi c \sqrt{\big(\frac{n_x}{L_x}\big)^2+\big(\frac{n_y}{L_y}\big)^2+\big(\frac{n_z}{L_z}\big)^2}$ where $(n_x, n_y, n_z) \in \mathbb{N}_0^3 \setminus (0, 0, 0)$ and approximate number of modes up to the cutoff frequency $f_c$ is given by: $\tilde{N}_{f_c} \approx \frac{4 \pi}{3}  V \big(\frac{f_c}{c}\big)^3$ where $V=L_xL_yL_z$.

Of special interest will be the basic axial resonant frequencies: $\omega_{[1,0,0]}=\frac{\pi c}{L_x}$, $\omega_{[0,1,0]}=\frac{\pi c}{L_y}$ and $\omega_{[0,0,1]}=\frac{\pi c}{L_z}$, because they will provide the data about the shape of the room.
\subsection{Reconstruction procedure}
Our goal is to reconstruct spatial periods of the wave vector grid from low-pass room impulse responses over each of the axes. The size of the room is assumed to be unknown and is 
\par

\algdef{SE}[DOWHILE]{Do}{doWhile}{\algorithmicdo}[1]{\algorithmicwhile\ #1}
\begin{algorithm}[H]
\begin{algorithmic}
\caption{ReSEMblE algorithm (Algorithm for the joint estimation of Room SizEs and ModEs)}
    \State \textbf{Input}: A set of measurements $\{m(x,y,z,t)\}_{i=1}^{M}$ at $M$ known locations $\mathbf{X}=[x,y,z]^T$ in space and T points in time. $\mathbf{R} \in \mathbb{C}^{T\times M}$ are measurements in matrix form and $\mathbf{r} \in \mathbb{C}^{TM}$ are measurements in a vectorized form. $f_p$ is frequency that separates data into 2 analysis procedures. \\
    \State \textbf{Output}: Estimated room size $\tilde{L}_x,\tilde{L}_y,\tilde{L}_z$ and \newline estimated room transfer function parameters: 
    \begin{itemize}
    \item expansion coefficients $\{\alpha\}_{n=1,v=1}^{N,V}$,
    \item resonant frequencies $\{\omega\}_{n=1}^{N}$ and damping $\{\xi\}_{n=1}^{N}$ 
    \item wave vectors $\{\mathbf{k}\}_{n=1,v=1}^{N,V}$ 
    \end{itemize}
    \\ $N$: number of modes, \\
    $V$: number of wave vectors per wave number. \\

    \Procedure{ReSEMblE}{$\mathbf{R}$, $\mathbf{X}$}
    \State Separate the measurements with $f_p$: $\mathbf{R}=\mathbf{R}^{l}+\mathbf{R}^{h}$.
    
    \For{$i_l \in \{1, ..., N_l \}$}
    \State \textbf{step 1}: estimate $(\omega_{i_l},\xi_{i_l})$ from $\mathbf{R}^{l}_{i_l}$
    \State \textbf{step 2}: estimate $\mathbf{k}_{i_l}$ from $\mathbf{r}^{l}_{i_l}$
    \State \textbf{step 3}: compute new residual $\mathbf{R}^{l}_{i_l+1}$
    \EndFor \\
    \State Recover the room size $\tilde{L}_x, \tilde{L}_y, \tilde{L}_z$ from basic axial room modes and form the regular wave vector grid.
    
    \For{$i_h \in \{N_l + 1, ..., N \}$}
    \State \textbf{step 1}: get $\omega_{i_h}$ and $\mathbf{k}_{i_h}$ from the wave vector grid
    \State \textbf{step 2}: estimate $\xi_{i_h}$ from $\mathbf{R}^{h}_{i_h}$
    \State \textbf{step 3}: compute new residual $\mathbf{R}^{h}_{i_h+1}$
    \EndFor
    
    \State Estimate the expansion coefficients $\{\alpha\}_{n=1,v=1}^{N,V}$ using least square approach.
    \EndProcedure
\end{algorithmic}
\end{algorithm}
\noindent
jointly estimated. All measured signals are separated into two components: low-pass $\mathbf{R}^{l}$ and high-pass $\mathbf{R}^{h}$. Analysis procedure is first applied to the low-pass component, which includes the estimation of the wave numbers and corresponding wave vectors. The bandwidth of this low-pass analysis is chosen in such a way that it covers reasonable sizes of rooms and removes the false modes that can appear below the first mode in RTF. With $f \in [20, 70] $Hz we cover room dimensions $L_x, L_y, L_z \in [2.45,8.575]$m for $c = 343 \frac{\textrm{m}}{\textrm{s}}$. This can easily be adjusted for rooms of unusual sizes.

\subsubsection{Estimation of \texorpdfstring{$\omega_{i_l}$}{TEXT}, \texorpdfstring{$\xi_{i_l}}{TEXT}$ and \texorpdfstring{$\xi_{i_h}}{TEXT}$}
In the \textit{low} part we define a unit-norm temporal dictionary with atoms of form: $\Theta[:,i]=\frac{\theta[i]}{\lVert \theta[i] \rVert}$, where $\theta[i]=e^{\xi_n[i]t}e^{j\omega_n[i] t}$ and $i$ is an index on a 2D grid of possible $(\omega_n, \xi_n)$, $\omega_n \in [0, \pi f_s]$ and $\xi_n \in [10\xi_0,0.1\xi_0]$, $\xi_0=-3\frac{\ln 10}{RT_{60}}$. The atoms with the highest correlation contains the solution pair. In the \textit{high} part the frequency is known, so we have only a 1D grid of possible values for the damping, which leads to a much simplified search.

\subsubsection{Estimation of \texorpdfstring{$\mathbf{k}_{i_l}}{TEXT}$}
The estimation of wave vectors is done with a structured group sparsity assumption - after estimating the wave number, we construct a sphere with a radius $\frac{\omega_n}{c}$ which follows from the assumption of lightly damped modes. 
We define a non-unit-norm spatio-temporal dictionary with atoms of form: $\Sigma[:,i]=e^{\xi_{i_l} t}e^{j\omega_{i_l} t}e^{\mathbf{X}\cdot \mathbf{k}[i]}$, where $\mathbf{k}$ are samples on this uniformly sampled sphere \cite{SphereSampling}.

On the surface of this sphere we search for a group of 8 wave vectors $[\pm k_x, \pm k_y, \pm k_z]^T$ which form a parallelepiped and which are aligned with the residual the most. In a case of tangential modes, the parallelepiped collapses over 1 dimension and shrinks to 4 wave vectors (e.g. $[\pm k_x, \pm k_y, 0]^T$), and axial modes are defined by 2 wave vectors (e.g. $[\pm k_x, 0, 0]^T$). 

In each iteration the best subgroup of 8 atoms has been estimated by applying a simultaneous version of matching pursuit (MP) \cite{MP} and the new residual is estimated by an orthogonal projection onto the space spanned by the union of all of the subgroups that were previously selected.

\section{RESULTS FOR RECONSTRUCTING THE K-SPACE OF A RECTANGULAR ROOM} 

In our solution we have relied on two types of structured sparsity expected in theory \cite{RoomAcoustics}: wave vector sparsity as nodes of parallelepiped and wave vector periodicity in the wave vector grid. How does this structured approach affect the data retrieval? As shown in \cite{Low, SpatioTemporal} efficient interpolation of the sound field is expected only within the part of the room surrounded by microphones used for training of the parameters.

We will present the performance of our approach on measurements made in a rectangular room with an approximate size $3\textrm{m} \times 5.6\textrm{m} \times 3.53\textrm{m}$. Properties of the chosen room are observed in \cite{RomainThesis}. Microphones are distributed randomly inside a $1\textrm{m}$ side cube in one half of the room and the sound source is in the other half of the room. Since we were processing real measurements, in order to have an idea about the approximate value of some of the parameters we want to estimate, we have applied the rational fraction polynomial curve fitting \cite{CurveFitting} based on the room mode shaped polynomials as basic fitting elements. This way we have retrieved approximate resonant frequencies and mode damping factors. During the curve fitting process, our wave numbers $k_n=\frac{\omega_n+j \xi_n}{c}$ appear in the poles of the fitted function \cite{EtienneThesis}:
$p_\omega(\mathbf{X})=\rho c^2 \omega q \sum_{n} \frac{\phi_n(\mathbf{X})\phi_n(\mathbf{X}_0)}{K_n[2 \xi_n \omega_n + j(\omega^2 - \omega_n^2)]}$.

\begin{figure}
\includegraphics[width=\columnwidth]{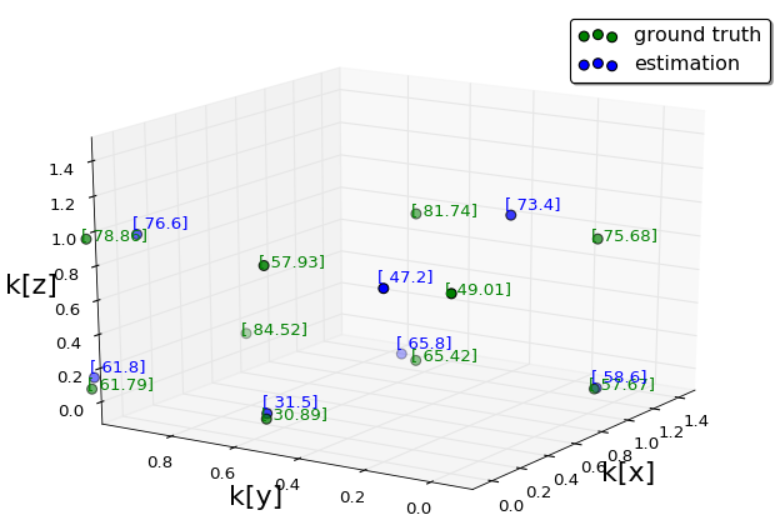}
\caption[width=\textwidth]
{ \label{fig:k-space} The estimation of wave vectors in $k$-space. The numbers next to the points indicate the corresponding eigenfrequencies (Hz). What we expect from theory in a case with perfectly rigid walls is plotted against the values we get from the measurements.}
\end{figure}

Figure \ref{fig:k-space}. shows the results for the estimation of the room mode resonant frequencies and their position in the $k$
-space in the \textit{low} part of the algorithm with 20 microphones. Here the $f_p$ frequency was set to be 70Hz. The basic axial modes are easily recognized and they give a fine approximation of the room size up to a few $\textrm{cm}$ away from ground truth. We can notice that the $k_x$ and $k_y$ component of the estimated wave vectors give a good approximation, but there is a slight deviation in the $k_z$ direction. This is attributed to the fact that in the room where the measurements were performed the floor is made from wood and ceiling is made of concrete. Also the slight deviation of the eigenfrequencies can be attributed to the fact that the search of the wave vectors was performed with a rigid wall model $\lVert \mathbf{k}_{q,r} \rVert = |k_q|$.

After applying the \textit{high} part of the algorithm, the Pearson correlation coefficient showed that the approximation is good (e.g. 82\% for only 19-microphone setting and $f_c=200\mathrm{Hz}$), but it should be further improved once the nature of the deviation of the wave vectors is efficiently characterized.

\section{CONCLUSION}
The proposed solution is suitable only for rectangular shaped rooms that are lightly damped, which was confirmed by the experiments. Also, the sound source has to be put in a position such that it excites all the axial modes. Although that the solution requires the $N$, $RT_{60}$ and $c$ parameters to be know, solution is not sensitive to their slight perturbation. The estimation of approximate structure of the $k$-space has lead to the reduction in the terms of number of required measurements and in the increase of the speed of the reconstruction without great losses of quality, but not for a broad range of frequencies. The higher we take the frequencies, the greater become the deviations. In the spirit of reproducible research, we have decided to open our \href{https://zenodo.org/record/1169161#.Wn3B-OjwY-U}{data} and \href{https://github.com/epfl-lts2/joint_estimation_of_room_geometry_and_modes}{code}.

\vfill

\def\UrlBreaks{\do\/\do-}
\label{sec:refs}
\bibliographystyle{IEEEbib}
\bibliography{main}
\end{document}